\documentclass{aa}

\usepackage{amsmath}
\usepackage{amssymb}
\usepackage{latexsym}
\usepackage{graphicx}
\usepackage[varg]{txfonts}
\usepackage{natbib}
\usepackage{color}
\bibpunct{(}{)}{;}{a}{}{,}

\newcommand{\hinode}{\textit{Hinode}\,}
\newcommand{\sot}{\textit{SOT}\,}
\newcommand{\xrt}{\textit{XRT}\,}

\begin{document}
\title{Magneto-seismological insights into the penumbral chromosphere and evidence for wave damping in spicules}

\author{R. J. Morton} 

\institute{Department of Mathematics \& Information Sciences, Northumbria University, Newcastle Upon Tyne,
NE1 8ST, UK}

\abstract{}
{The observation of propagating magneto-hydrodynamic kink waves in magnetic structures and measurement of their properties (amplitude, phase 
speed) can be used to diagnose the plasma conditions in the neighbourhood of the magnetic structure via magneto-seismology. We aim to reveal 
properties of the chromosphere/Transition Region above the sunspot penumbra using this technique. }
{\hinode \sot observed a sunspot as it was crossing over the limb, providing a unique side on view of the atmosphere above a sunspot. The presence of 
large spicule-like jets is evident in \ion{Ca}{II} H images. The jets are found to support transverse wave motions that displace the central axis of the 
spicules, which 
can be interpreted as the kink wave. The properties of a specific wave event are measured and used to determine the magnetic and density stratification 
along the structure. In addition, we measure the width of the spicule and the intensity profile along the structure in order to provide a test for the 
magneto-seismological results.}
{The measurements of the wave properties reveal an initial rapid increase in amplitude with height above the solar surface, followed by a decrease in 
amplitude. The magneto-seismological inversions suggests this initial increase corresponds to large changes in density and magnetic field strength. In 
addition, we provide the first measurements of spicule width with height, which confirm that the spicule under goes rapid expansion. The measured 
rates of expansion show good agreement with the results from the magneto-seismology. The observed rapid variations in plasma parameters is 
suggested to be partly due to the presence of a gravitational stratified, ambient atmosphere. Combining width measurements with phase speed 
measurements implies the observed decrease in wave amplitude at greater heights can be explained by wave damping. Hence, we provide the first 
direct evidence of wave damping in chromospheric spicules and the quality factor of the damping is found to be significantly smaller than estimated 
coronal values.}
{}

\keywords{Sun: Chromosphere, Waves, magnetohydrodynamics (MHD), Sun: sunspots, Sun:oscillations}

\date{Received /Accepted}

\titlerunning{Active region seismology}
\authorrunning{Morton}

\maketitle

\section{Introduction}
The observation of magnetohydrodynamic (MHD) wave phenomenon in the solar atmosphere has become a common occurrence in the last two decades 
(see, e.g. \citealp{BANetal2007}, \citealp{DEM2009}, \citealp{MATetal2013} for reviews). The observations have established that waves are ubiquitous in 
many different structures, e.g. coronal loops (\citealp{TOMetal2007}), spicules (\citealp{DEPetal2007}), fibrils (\citealp{MORetal2012c}), prominences 
(\citealp{HILetal2013}). Aside from their potential importance for heating the atmospheric plasma, the waves can also be exploited for magneto-
seismology (\citealp{NAKVER2005}, \citealp{DEMNAK2012}). 

Magneto-seismology is still a relatively fledging field of solar research and is growing in popularity, applicability, and examples (\citealp{UCHIDA1970}; 
\citealp{ROBetal1984}; \citealp{BANetal2007}). The basic principles are based on the assumption that MHD waves are guided by localised magnetic 
structures in the solar atmosphere, with imaging observations appearing to support the existence of such features, for example coronal loops (e.g. 
\citealp{ASCNIG2005}) or chromospheric fibrils/spicules (\citealp{RUT2006}). The confinement of certain wave modes to the magnetic structure means 
that the wave properties (e.g. period, amplitude, phase speed) are determined by the local plasma parameters, i.e. those of the supporting structure and 
the local ambient plasma. Hence, measurement of the wave properties and their inversion can reveal information about the 
conditions in the neighbourhood of the observed wave, typically information that is difficult to measure from radiation, be that emission measure, 
polarisation, etc.

To date, there has been significant effort focused on the theoretical development of the field (\citealp{ANDetal2009}; \citealp{RUDERD2009}; 
\citealp{DEM2009}), inspired by the \textit{TRACE} observations of oscillating coronal loops (\citealp{NAKetal1999}; \citealp{ASCetal1999}) and 
propagating intensity disturbances (\citealp{DEMetal2002}). A wave mode that has received significant interest is the fast kink mode, characterised by 
the physical displacement of the supporting wave guide in the direction perpendicular to the magnetic field (\citealp{SPR1982}; \citealp{EDWROB1983}). 
However, the early observational examples of the kink wave mode were related to large amplitude, standing modes excited by flare blast waves. 
Although their potential for magneto-seismology has been demonstrated (e.g. \citealp{NAKOFM2001}; \citealp{ARRetal2007}; \citealp{VERERDJES2008}; 
\citealp{GOOetal2008}; \citealp{VERetal2010b}; \citealp{VERetal2013b}), the large wavelengths associated with standing modes (i.e. two times the 
typical coronal loop length) leads to additional complexities, e.g. consideration of the curvature of the coronal loop (\citealp{VANetal2004}; 
\citealp{VANetal2009b}; \citealp{RUD2009}) and non-planarity (\citealp{RUDSCO2011}). Further, the loop also spans a number of density and magnetic 
scale heights which can lead to variations in the cross-sectional area (e.g. \citealp{RUD2009}; \citealp{ERDMOR2009}).

More recently, the advent of high-resolution spectral and imaging instrumentation (e.g. \textit{Hinode Solar Optical Telescope (SOT)}, \textit{ROSA} at 
the Dunn Solar Telescope, \textit{CRiSP} at the Swedish Solar Telescope) has led to the observations of ubiquitous propagating kink waves in the 
chromosphere (\citealp{HEetal2009, HEetal2009b}; \citealp{PIEetal2011}; \citealp{OKADEP2011}; \citealp{KURetal2012}; \citealp{MORetal2012c}). In 
spite of the vast amount of observed propagating wave phenomenon and the many data sets (e.g. the numerous \ion{Ca}{II} limb data sets), very little 
use has been made of this rich vein of potential information for magneto-seismology. This opportunity has not escaped all, for example, 
\cite{VERTetal2011} provide the first magneto-seismological diagnostic of the quiet chromosphere using \sot \ion{Ca}{II} observations, obtaining 
estimates for the density and magnetic field gradients and the ionization fraction with height. Further, \cite{KURetal2013} has applied similar techniques 
to observations of kink waves in chromospheric mottles. The volume of data and opportunity for the study of the chromosphere and Transition Region 
via magneto-seismology will no doubt be increased by the high-resolution data from the \textit{Interface Regions Imaging Spectrograph}.

The observation and corresponding magneto-seismology of both standing and propagating waves is not trivial. Typically, significant simplifying 
assumptions are made about the structure that supports the observed wave during the derivation of theory and estimated values for parameters are 
used as inputs for the seismological inversion (e.g. density contrasts between internal and ambient plasmas). It is then not clear how accurate the 
derived magneto-seismological estimates of the various plasma parameters are, relying only on previous estimates of the parameters as guidance.  On 
this note, \cite{VERetal2013} has demonstrated the correspondence between values of magnetic field strength derived from magneto-seismology and 
from magnetic field extrapolations of the same coronal loop. While such a comparison can be achieved for coronal features, the modelling of 
chromospheric magnetic fields from magnetograms is still an uncertain and ongoing process due to the non-potentiality of the chromospheric fields 
(\citealp{JINetal2011}; \citealp{MEYetal2013}). Further, imprudent interpretations of magneto-seismological results can lead to erroneous conclusions 
about local plasma conditions. For example, failure to consider the influence of wave damping when exploiting the amplitudes of propagating waves for 
magneto-seismology will undoubtedly lead to underestimates of density and magnetic field gradients.

\bigskip
Here we present a unique observation of wave phenomena in a chromospheric plasma that provides a test of the theoretically derived relations that have 
been used previously for magneto-seismology exploiting propagating waves (\citealp{VERTetal2011}; \citealp{MORetal2012}; \citealp{KURetal2013}).  
The waves are observed in spicule-like jets that exist in the penumbral region of a sunspot. The jets reach much higher into the solar 
atmosphere than the surrounding spicules associated with the plage, suggesting that the jets follow the strong magnetic field above the sunspot. In 
addition, the almost vertical magnetic field lines associated with the sunspot means that the spicule-like features are easier to observe and study in 
comparison to their plage counterparts, minimising the typical confusion associated with inclined spicules crossing each other. For further details on the 
jets and the active region, see \cite{MOR2012}.

The spicules demonstrate evidence for oscillatory motions which we are able to measure, obtaining amplitudes and phase speeds as a function of 
height. The magnetic field gradient along the spicule is estimated via magneto-seismology, along with the plasma gradient. In addition, there is also a 
measurable change in the width of the spicule with height, which enables an independent test for the seismologically estimated magnetic field gradient. 
It is demonstrated that the measured wave amplitude is likely subject to the effects of damping, hence, we provide the first direct evidence of damped 
kink waves in the chromosphere. The quality factor of the damping is found to be much smaller than those previously measured in the corona, implying 
that there is an enhanced damping of waves in the chromosphere and supporting indirect measurements of enhanced damping in the 
chromosphere/transition region (\citealp{MORetal2013b}).

In addition, this result highlights that if the measured amplitude is solely relied upon for magneto-seismological inversion, it would provide incorrect 
information about the local plasma conditions. Using a combination of the measured width and phase speed, a more realistic estimate for the variation 
in local plasma parameters is obtained. In general, the following investigation provides the first seismological estimates for the plasma conditions in the 
upper penumbral chromosphere. 

\begin{figure}[!tp]
\centering
\includegraphics[scale=1.1, clip=true, viewport=0.cm 0.0cm 8.2cm 7.8cm]{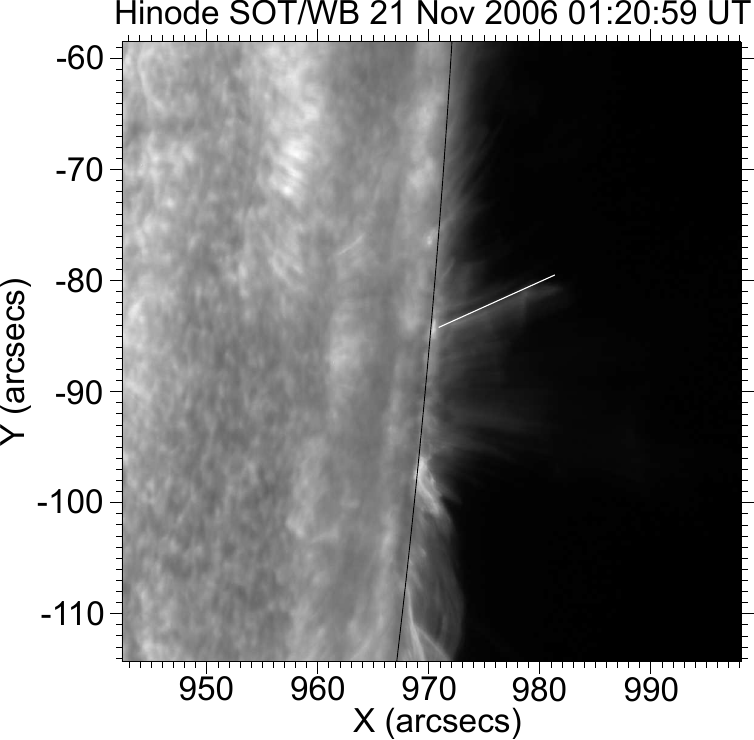} 
\caption{\hinode \sot field of view. The image also shows the position of the \ion{Ca}{II} limb (black solid line), and the
central axis of the spicule (solid white). Cross-cuts are taken perpendicular to this central axis  }\label{fig:fov}
\end{figure}

\section{Observations and data reduction}
The data were obtained by \textit{Hinode Solar Optical Telescope} (\sot -
\citealp{SUEetal2008}) at {01:11:00}~UT 21 November 2006 till
{02:00:00}~UT on the same day. \hinode/\sot viewed a region on the
west limb with the \ion{Ca}{II} H broadband filter, which has a pixel size
of $0.054''$ but is diffraction limited to $0.2''$ ($145$~km). The
cadence of the data is $8$~s. We performed the usual processing
routine for \sot data sets with \textit{fg\_prep.pro}. However, the
data still possess a significant drift over time and jitter. The drift is 
first corrected for by tracking the limb over the time series and removing the
trend. Next, the remaining jitter is removed by careful alignment of the data using
cross-correlation. The remaining frame to frame movement is found to have an RMS value of 0.1 pixels.

In the time series, \hinode {was} following the active region 10923.
Hinode tracked its progress across the disk and the active region
has been the subject of a number investigations (e.g.
\citealp{KATetal2007}; \citealp{TIAetal2009}). In the data series used here, the sunspot is
about to cross the limb (Fig.~\ref{fig:fov}) allowing a novel
side-on view of the sunspots structure. The images clearly show
vertical and inclined magnetic fields emanating from the region into
the lower corona (\citealp{MOR2012}). In the following, we pay particular attention to one of these spicule or spicule-like jets, which
is highlighted in Fig.~\ref{fig:fov}. The spicule is found to support a MHD fast kink wave, which will be the focus of the 
following investigation. 

\begin{figure*}[!tp]
\centering
\includegraphics[scale=1.1, clip=true, viewport=0.cm 0.0cm 12.2cm 16.8cm]{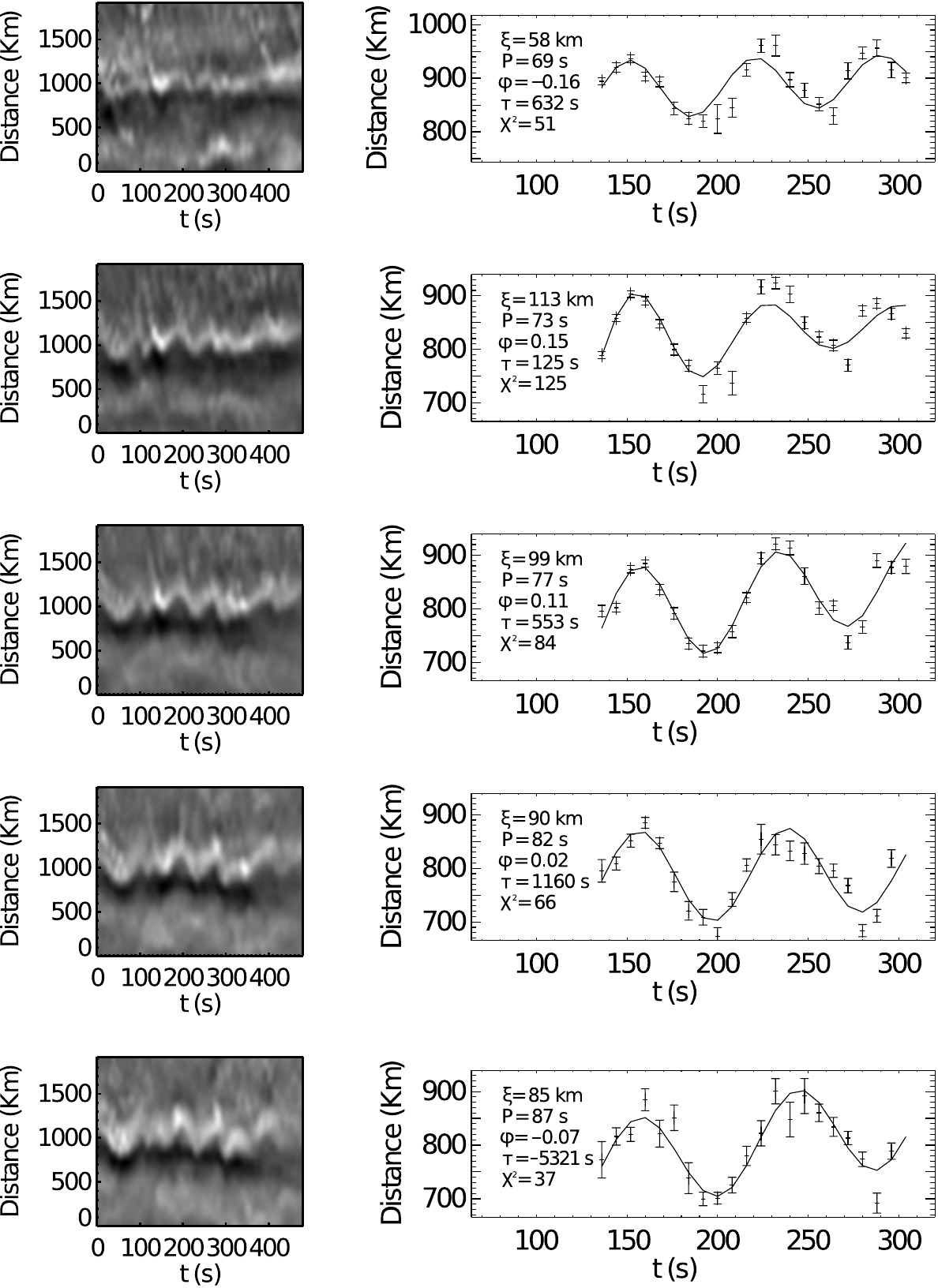} 
\caption{The observed kink motion. The left hand column panel shows time-distance diagrams from various cross-cuts. Each cross-cut shown is 
separated by 800~km, with the lowest cross-cut on the top row. The right hand column shows the results of the Gaussian fits (crosses) with the 
$\sigma$ error bars on the measured position. The solid line is the result of the damped sinusoidal fit. The given distance corresponds to the position 
along the cross-cut and the time is seconds from {01:20:51}~UT.}\label{fig:tds}
\end{figure*}

\begin{figure*}[!tp]
\centering
\includegraphics[scale=0.6, clip=true, viewport=0.0cm 0.0cm 10.cm 10.cm]{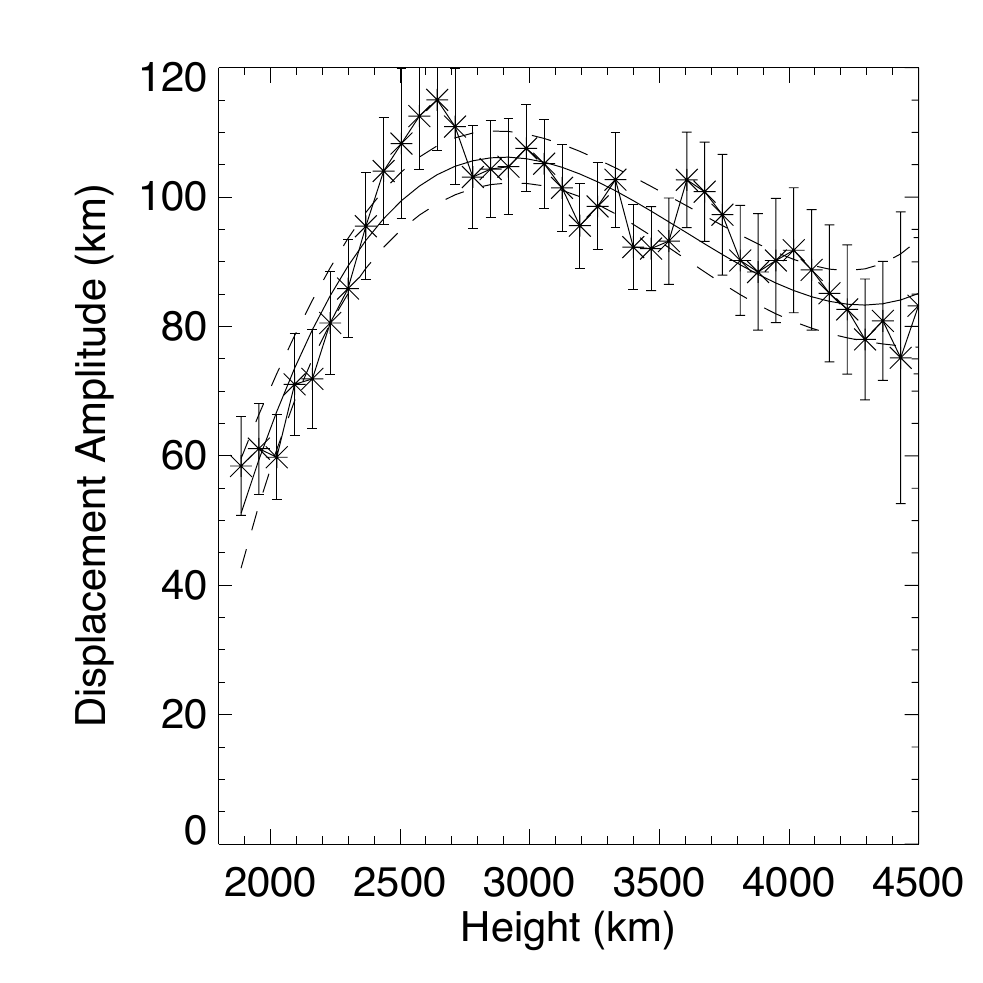} 
\includegraphics[scale=0.6, clip=true, viewport=0.0cm 0.0cm 20.cm 10.cm]{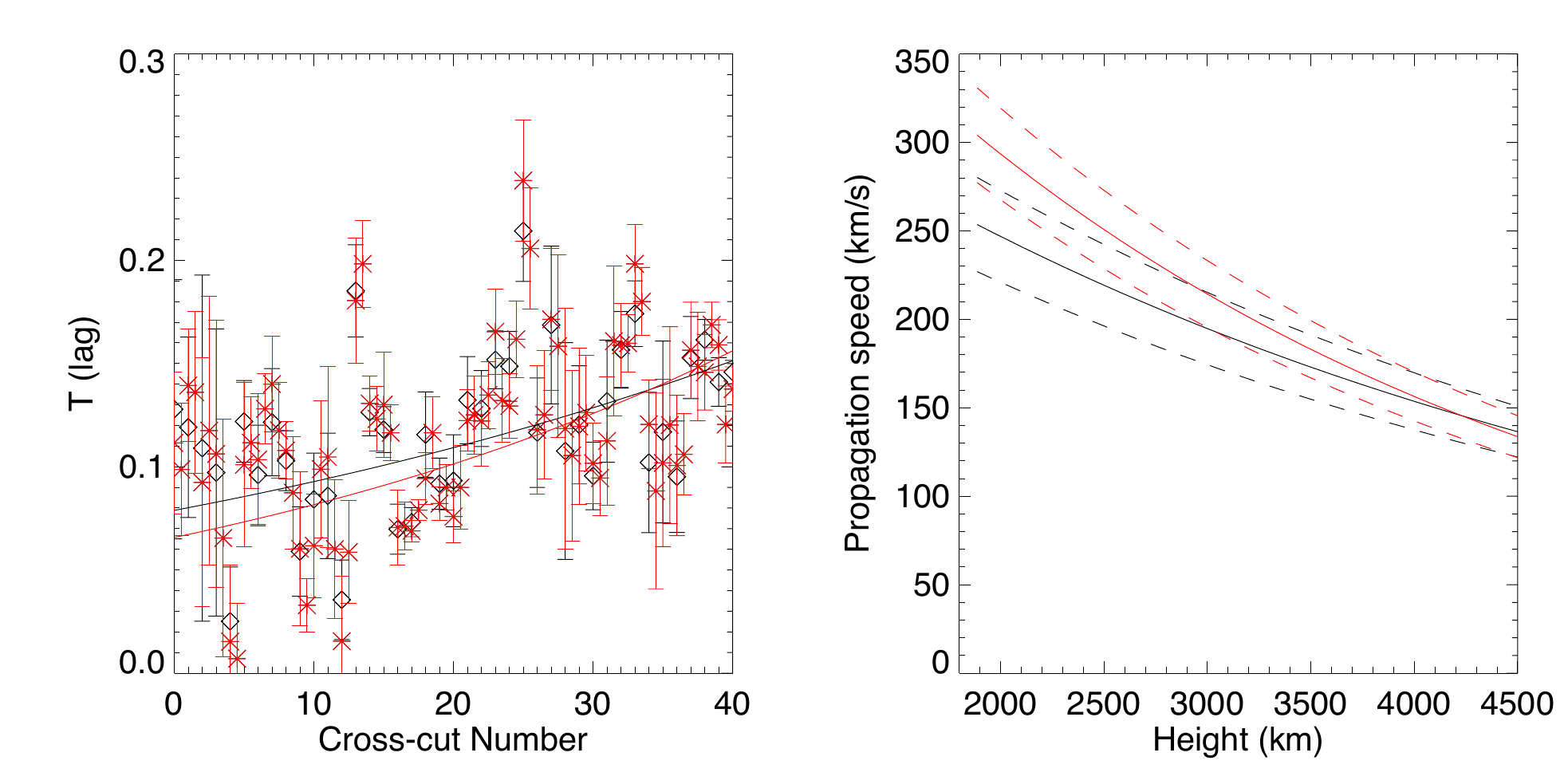} 
\caption{Measured wave properties. The left panel displays the measured displacement amplitude (stars) as a function of height in the atmosphere 
relative to $\tau_{5000}=1$. The vertical bars shows the $\sigma$ error on the measured quantity. The solid black line corresponds to the cubic fit to 
the amplitude profile and the dashed lines show the $2\sigma\, (95\%)$ confidence level of the fit. The middle panel displays the lag value measured 
from cross-correlation of neighbouring time-distance diagrams separated by 40~km (red stars) and 80~km (black diamonds). The red and black solid 
lines are the corresponding exponential fits to measured lags. The right hand panel displays the phase speed calculated from the fits to the lag data 
points. The red and black dashed lines correspond to the $\sigma$ confidence levels for the phase speeds.}\label{fig:observ}
\end{figure*}

\section{Methods}
The techniques used to measure the wave and spicule properties are now described, an account of
sources of error and error propagation is also given and a brief overview of the relevant magneto-seismological theory is provided.

\subsection{Wave fitting}\label{sec:wavefit}
Following previous publications (\citealp{MORetal2012c, MORetal2013}; \citealp{MORMCL2013}), time-distance diagrams are used to analyse the 
observed wave motion. An unsharp mask (USM) procedure is applied to each frame in order to enhance the spicule compared to the background 
emission. A time-averaged image of the spicule is used to determine the main axis of the structure and we note the angle between the jet and the 
normal to the surface is 28.6~degrees. Each time frame is interpolated (cubic) to a grid a factor of $6\times6$ greater than the original and forty cross-
cuts are placed perpendicular to the main axis of the spicule, each separated by $\sim78$~km. Once the cross-cuts are determined, the results are re-
scaled to the original spatial sampling and the time-distance diagrams are created. The interpolation step is used to help reduce any differences 
between each cross-cut position and separations that may arise during the mapping of intensities due to the discrete spatial sampling. Cross-cuts are 
also generated in the same manner from the non-USM data, which are used for error calculation. Examples of the generated time-distance diagrams are 
shown in Fig.~\ref{fig:tds}. 

To fit the wave observed in the time-distance diagram, a Gaussian function is fit to the cross-sectional flux profile in each time-frame. The central 
position of the fitted Gaussian is taken to represent the centre of the spicule. As noted in \cite{MORMCL2013}, to accurately determine the errors
on the measured central position of the spicule, the fitting function has to supplied with the associated errors in the values of intensity. In 
order to assess the uncertainty for \sot, the procedure laid out in \cite{ASCetal2000} and \cite{YUANAK2012} is followed. For each flux value, $F$, 
associated with a pixel, we combine the uncertainties in units of data number (DN). The errors arise from photon Poisson noise, dark current noise, flat-
field noise, electronic readout noise, compression noise, and subtraction noise. The total error is given by
\begin{eqnarray}
\sigma^2_{noise}(F)&=&\sigma^2_{photon}(F)+\sigma^2_{dark}+\sigma^2_{flat}+\sigma^2_{readout}\\
&&+\sigma^2_{compress}+\sigma^2_{subtract}.
\end{eqnarray} 
The photon noise is dependent upon the gain, $G$, of the CCD and the flux of photons received. \cite{SHIMetal2007} gave the value of the gain for the 
BFI as $G=64$~e/DN, and the photon noise is given by the usual relation $\sigma^2_{photon}(F)=F/G$.

In order to assess the dark current and flat field noise, the dark and flat field files were used to generate histograms of the dark/flat values. The CCD is 
composed of two separate sections with different dark currents and flat field values. A Gaussian was fit to the distribution generated from each section. 
The standard deviation of the two dark and flat field values are 0.21~DN, 0.298~DN and 0.16~DN, 0.16~DN, respectively. In order not to underestimate 
the errors, the largest value for the dark current is taken.

To estimate readout noise, \textit{fg\_prep} was run without readout calibration performed. This was subtracted from the fully calibrated data and the 
residual  values were plotted in a histogram and a Gaussian fit gave the standard deviation as 0.51~DN.

The \hinode data is compressed with a 12bit \textit{jpeg} data compression algorithm using Q table 95. \cite{KOBetal2013} suggest that the 
\textit{jpeg} compression used on \hinode \xrt images introduces uncertainties of 1.5~DN. Similar ranges of DN are obtained for \xrt and \sot 
images suggesting compression would provide similar errors. To avoid underestimation of the error we set the compression noise at 2~DN. 

Digitisation of the signal from the analog-to-digital converter provides three lots of 0.5~DN uncertainty (raw, darks and flats), hence, 
$\sigma_{subtract}=\sqrt{0.5\times3}$.

\noindent The total error is then given by
\begin{equation} 
\sigma^2_{noise}(F)=\frac{F}{64}+5.1,
\end{equation}
and the error on intensity values are then calculated for each USM time-distance diagram from the corresponding non-USM cross-cut. Once calculated, 
the errors are supplied to the Gaussian fitting routine to provide the error on the position of central axis of spicule.

The error on the central position is then summed with the uncertainty from the alignment procedure ($0.1$~pixels) to give the total error on
the determined central position of the spicule. On doing this for each time slice in a time-distance diagram, we obtain a series of
data points related to the motion of the spicule in time (Fig.~\ref{fig:tds}). We fit these data points with a sinusoidal function multiplied by an
exponential, namely
\begin{equation}
G(t)=\xi\sin\left(\frac{2\pi t}{P}-\phi\right)\exp\left(-\frac{t}{\tau}\right),
\end{equation} 
where $\xi$ is the transverse displacement, $P$ is the period of the oscillation, $\phi$ is the phase, and $\tau$ is the damping/amplification
term depending upon the sign of $\tau$. The measured amplitude in each time-distance diagram is plotted in Fig.~\ref{fig:observ}

\subsection{Spicule width measurements}\label{sec:wid}

\begin{figure}[!tp]
\centering
\includegraphics[scale=0.5, clip=true, viewport=0.cm 0.0cm 16.cm 14.6cm]{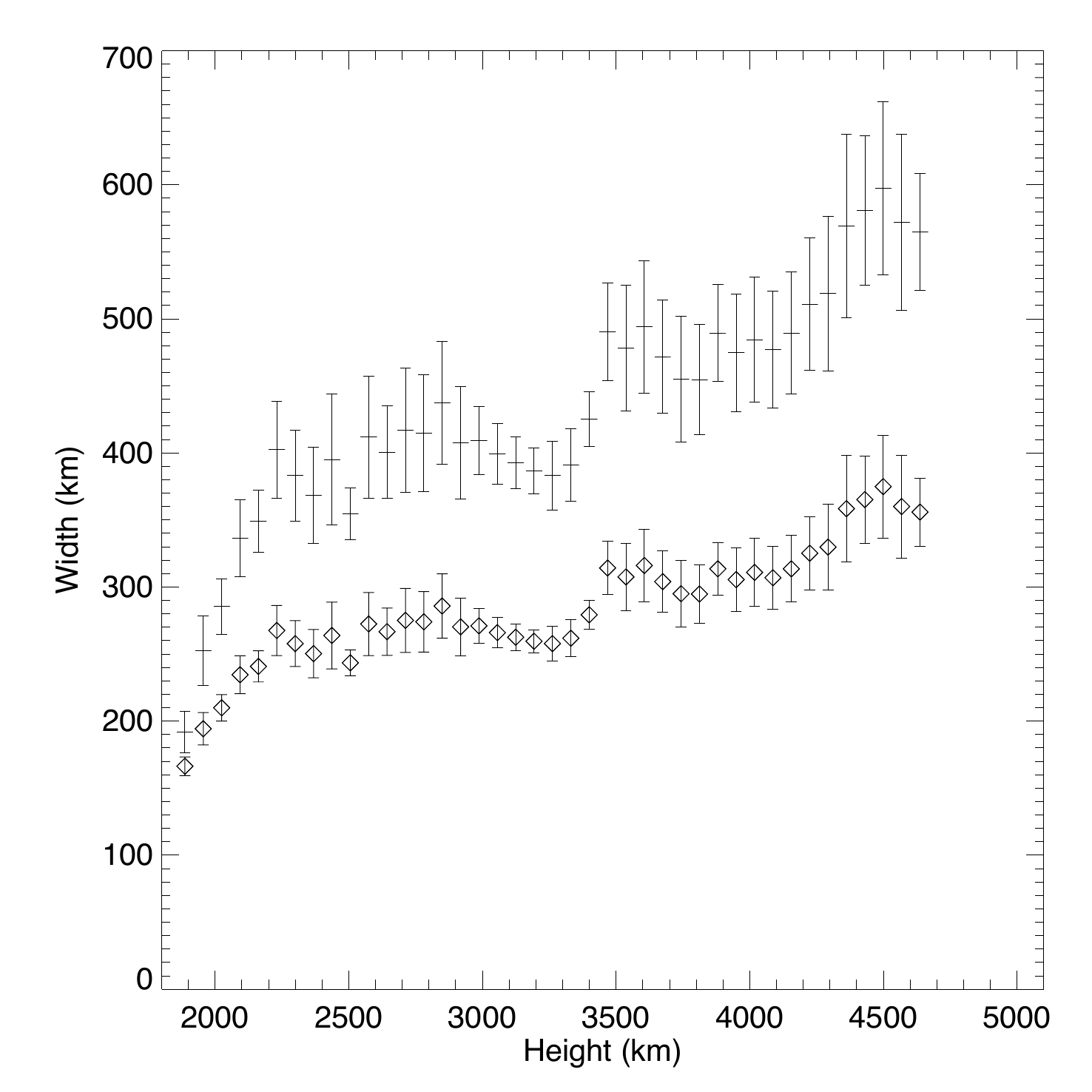} 
\caption{Measured and {estimated} spicule widths as a function of height in the atmosphere 
relative to $\tau_{5000}=1$. The diamond data points are the spicule widths determined form the fitting of a Gaussian to the 
cross-sectional flux profile. The crosses are the estimates for the physical diameter of the spicule calculated from the modelling (referred to as the 
{estimated} width). The error bars correspond to the standard error. }\label{fig:width_obs}
\end{figure}

In order to provide a comparison for the magneto-seismological results (Section~\ref{sec:mag_seis}), the width of the spicule is measured at different 
heights. Similar to the wave fitting, a Gaussian plus a linear function is fit to the cross-sectional flux profile in each time slice in a time-distance 
diagram. A measure of the half-width of the spicule is given by the $\sigma$ value of the Gaussian. Averaging the $\sigma$ values over time and 
taking the standard error of the mean provides the data points and error bars in Fig.~\ref{fig:width_obs}.

It is well known that the measured width of solar structures in images is, however, not the actual radius of the structure (\citealp{WATKLI2000}; 
\citealp{DEF2007}; \citealp{BROetal2013}). The ability to determine the actual radius is impeded by the finite spatial resolution and the 
Point Spread Function (PSF) of the telescope. In addition, the observed radiation from optically thick lines, such as Ca II H, is affected by the geometry of 
the structure in which the radiation originates, adding further complications. To assess how some of these limitations affect the measured radius, we 
provide an estimate of the influence of resolution and PSF on the difference between actual and measured radius, following a similar approach to 
\cite{WATKLI2000}. 

First, a monolithic structure of length $L$ with circular cross-section of radius $r$ is defined on a $1000\times1000\times1000$ grid with each grid 
cube representing $4\times4\times4$~km$^3$. Each grid cube inside the structure is given the same value on an arbitrary intensity scale. The values 
are then summed in a direction perpendicular to the direction of the central axis providing a two-dimensional image of the structure. This method 
implicitly assumes that the plasma is optically thin and radiation from each volume element can be summed to give the line-of-sight intensity. 

The PSF for \sot BFI is particularly complicated and has only been determined previously for particular values of wavelength (\citealp{WED2008}, 
\citealp{MATetal2009}). The ideal PSF for G-band is given by \cite{SUEetal2008} as 0.2'', and \cite{WED2008} noted that the ideal PSF varies with the 
inverse of wavelength. Using this, the ideal PSF for 396.85~nm is calculated as $\sim$0.166''.

The 2-D image of the structure is then interpolated to a grid with grid points equal in size to the \sot pixel size, i.e. $39.1\times39.1$~km$^2$. This 
image is then convolved with an Airy function of width $3.07$ pixels in an order to replicate the PSF. The width of the structure in this new image is 
then fitted with a Gaussian and the $\sigma$ measured. Doing this for a number of different radii, we build up a series of data points providing the 
difference between measured $\sigma$ and actual radii. Fitting the data points with a quadratic provides a relation between the two quantities which we 
use to determine an estimate for the radius.   

It is clear that this technique is subject to many idealising assumptions, however, a more thorough and accurate method would require advanced 
models of \ion{Ca}{II} line formation and a complete determination of the PSF at 396.85~nm. Such additions are outside the scope of this work. The 
current results should be taken as a guide to the expected influence of the PSF on the measured radius and the `actual' radius of the spicule. We refer 
to the quantity obtained via this technique as the {estimated} radius (width).

\subsection{Propagation speed determination}\label{sec:phase}

The propagation speed of the wave is also determined by exploiting the time-distance diagrams. The basis of the technique relies on 
cross-correlation of neighbouring time-distance diagrams. The apparent shift of the wave motion in the time direction provides the signal lag between 
the two images, for which the propagation speed is then given by
\begin{equation}\label{eq:lag}
c_{p}=\frac{dx}{dtT},
\end{equation}
where $dx$ is the distance between the cross-cuts used to generate the time-distance diagrams, $dt$ is the time difference between each pixel, and
$T$ is the lag value.

In order to improve the accuracy of the phase speed determination, the time-distance diagrams are first interpolated so that they are mapped onto a 
grid twice the size of the original, hence each pixel has dimensions 19.6~km by 4~s. The uncertainty on each measurement from the cross-correlation 
is unknown, although it is believed to be $<0.1$~pixel. The second step is the application of cross-correlation for a sliding window in the 
time direction giving multiple measurements of the lag time. The lag value is calculated as the average value of the measurements and the error is taken 
as the standard error of the results.

In Fig.~\ref{fig:observ}, the lag values calculated for the time-distance diagrams separated by $\sim80$~km are displayed. In addition, time-distance 
diagrams are generated for additional cross-cuts placed mid-way between the previous cross-cuts. This provides 80 cross-cuts separated by 
$\sim40$~km. Again, these are interpolated and the lag is calculated. Multiplying these additional values by two we obtain lags of similar magnitude to 
those found for the $\sim80$~km separated cross-cuts (see Fig.~\ref{fig:observ}). This provides confidence that the obtained lag values are genuine.

The values of measured lags are close to 0.1~pixels, with standard deviations for the $\sim80$~km lags ranging between 0.006-0.08~pixels, with an 
average of 0.03~pixels. For the $\sim40$~km lags, the measured lags are $\sim$0.05~pixels with standard deviations ranging between 
0.003-0.03~pixels, with an average of 0.01~pixels.

Next, a weighted fit of an exponential function to the lag data points is performed. A fit of a linear function was also performed but the exponential 
function was found to have a smaller $\chi^2$ value. The residual values to the fit have root-mean-square amplitudes of 0.035 and 0.021 for the 
$\sim80$~km and $\sim40$~km separations, respectively. Both of these values are larger than the average standard deviations, which indicates that 
some systematic error is present in the analysis. It is difficult to determine exactly where the systematic errors enter, one option is this occurs during 
the determination of cross-cut position, i.e. an uncertainty in the relative position of each cross-cut. This error could possibly be reduced if the data is 
interpolated to a larger grid when initially creating the cross-cuts and time-distance diagrams. 
 
Bearing this in mind, the propagation speed and associated errors are determined using Eq.~(\ref{eq:lag}), with $T$ given by the fit and uncertainties 
are the $\sigma$ errors to the fit, $dt$ is $4\pm0.05$~s and $dx=80\pm7$~km or $dx=40\pm7$~km. The error in $dt$ is given by the accuracy of 
the clock on-board \hinode, which gives times rounded to $0.1$~s. The uncertainty in $dx$ is taken from knowing the position of the cross-cut to 1/6 
of a pixel, i.e. from the initial interpolation. It is found that the propagation speeds determined for both sets of lags largely agree within $\sigma$ 
errors (Fig.~\ref{fig:observ}).  

\begin{figure}[!tp]
\centering
\includegraphics[scale=0.5, clip=true, viewport=0.cm 0.0cm 11.cm 23.6cm]{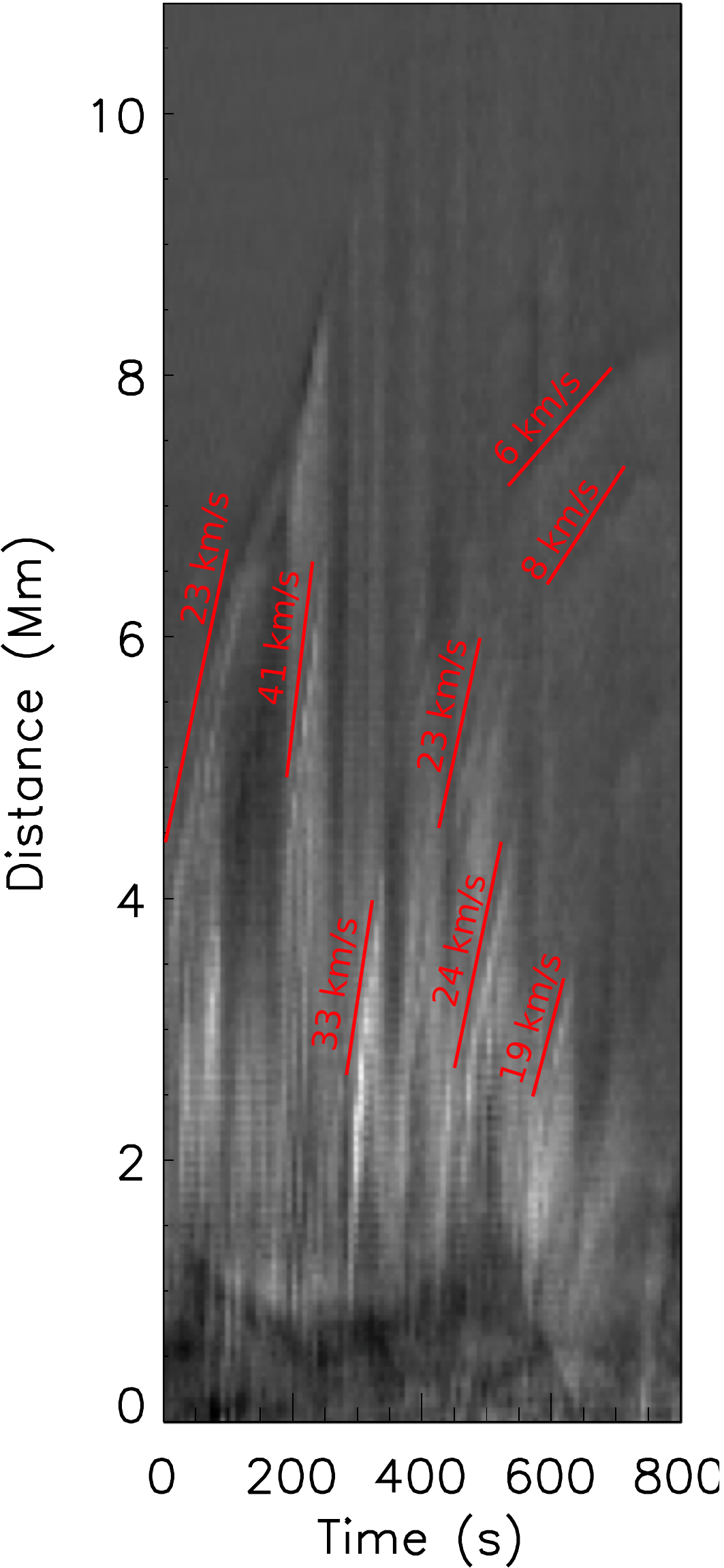} 
\caption{Flows along the spicule. The image shows a time-distance diagram generated from the unsharp masked data and taken parallel to the spicule 
axis. The straight lines highlight the measured up-flows and the measured flow speeds are given.
The time is shown in seconds from {01:18:11}~UT.}\label{fig:flow_obs}
\end{figure}

\subsection{Theoretical considerations}\label{sec:theory}
Having measured the amplitude and propagation speed as a function of distance along the spicule, these quantities can be used
to estimate gradients in magnetic field and plasma density. The WKB approximation for the wave equation 
governing a kink wave propagating along a stratified flux tube (\citealp{VERERD2008}; \citealp{RUDetal2008}; \citealp{MORetal2012}) gives the solution,
\begin{equation}\label{eq:wkb}
\xi(z)=C\sqrt{\frac{c_k(z)}{\omega}}R(z),
\end{equation}
where
\begin{equation}\label{eq:phase}
c_k(z)=\frac{\langle B(z)\rangle}{\sqrt{\mu_0\langle\rho(z)\rangle}}, \qquad \langle B(z)\rangle\propto\frac{1}{R(z)^2}.
\end{equation}
Here, $c_k(z)$ is kink phase speed, $C$ is some constant, $\omega=2\pi/P$ is the angular frequency of the wave, $R(z)$ is the radius of the flux tube, 
$\langle B\rangle$ is the local average magnetic field strength, $\langle\rho\rangle$ is the local average density, and $\mu_0$ is the magnetic 
permeability. A similar technique is used in \cite{VERTetal2011}, although they solve the governing equation under the assumption the phase speed 
changes exponentially with respect to z. {Note that these equations are derived in the thin tube approximation, i.e. where the wavelength is much 
longer than the radius of the wave-guide. The wavelength, $\lambda$, is given by the product of the period and the phase speed of the wave. From 
Figures~\ref{fig:tds} and \ref{fig:observ}, the wavelength can be calculated to be on the order of $10,000$~km, hence, the ratio of 
$r/\lambda\sim0.03$ (using the measurements in Figure~\ref{fig:width_obs}) and the thin tube approximation is satisfied for the current observation. }

Equation~(\ref{eq:wkb}) gives the expected amplitude evolution for a propagating wave if it is assumed that the influence of wave damping mechanisms 
can be neglected. However, there is evidence indicating that wave damping occurs in observations of waves in solar plasmas (e.g. 
\citealp{ASCetal1999}, \citealp{TOMMCI2009}). It is generally considered that resonant absorption is responsible for the observed damping, with a good 
agreement between theory and observations (\citealp{RUDROB2002}, \citealp{VERetal2010}, \citealp{DEMPAS2012}  \citealp{TERetal2010c}; 
\citealp{SOLetal2011,SOLetal2013}). \cite{SOLetal2011} derived the function that describes the evolution of the amplitude for propagating kink waves 
with the inclusion of resonant absorption. Consideration of their Eq.~(38) suggests that Eq~(\ref{eq:wkb}) would be modified as follows, 
\begin{equation}\label{eq:wkb_res}
\xi(z)=C\sqrt{\frac{c_k(z)}{\omega}}R(z)\exp\left(-\int_0^z\frac{1}{L_D(z)}dz\right),
\end{equation} 
where the damping length is
\begin{equation}\label{eq:damp_l}
L_D(z)=\frac{c_k(z)\xi_E}{f},
\end{equation} 
and $\xi_E$ is the quality factor and $f=1/P$ is the frequency of the wave. 

\smallskip
For now, it is assumed that damping can be neglected and magneto-seismological relations useful for inversions of the
measured wave properties are derived. Upon dividing Eq.~(\ref{eq:wkb}) by itself evaluated at $z=0$ and rearranging, the following is obtained
\begin{equation}\label{eq:rad}
\frac{R(z)}{R(0)}=\sqrt{\frac{c_k(0)}{c_k(z)}}\frac{\xi(z)}{\xi(0)}.
\end{equation}
This relation allows the estimation of the magnetic expansion along the flux tube using the measured values of amplitude
and propagation speed.

To calculate the uncertainty in $R$, Eq.~(\ref{eq:mag}) is written in terms of the independently measured quantities, namely
 \begin{equation}\label{eq:rad2}
\frac{R(z)}{R(0)}=\sqrt{\frac{T(z)}{T(0)}}\frac{\xi(z)}{\xi(0)},
\end{equation}
where $c_k$ is taken to equal $c_p$, and following standard rules for error propagation,
\begin{equation}\label{eq:rad_er}
\delta R = \frac{R(0)}{\sqrt{T(0)}\xi(0)}\sqrt{\frac{\xi^2\delta T^2}{4T}+T\delta\xi^2},
\end{equation}
where $\delta y$ is the $\sigma$ error associated with the quantity $y$. Similar equation can be derived for the magnetic field, $B$,
 \begin{equation}\label{eq:mag}
\frac{<B(z)>}{<B(0)>}={\frac{T(0)}{T(z)}}\frac{\xi(0)^2}{\xi(z)^2},
\end{equation}
with the associated uncertainty
\begin{equation}\label{eq:mag_er}
\delta B = <B>(0)T(0)\xi(0)^2\sqrt{\frac{\delta T^2}{T^{4}\xi^4}+\frac{4\delta\xi^2}{T^2\xi^6}}.
\end{equation}

\noindent Further, the gradient of the density can be obtained by combing Eqs.~(\ref{eq:wkb})-(\ref{eq:phase}), where 
it is easily found that
\begin{equation}\label{eq:den}
\frac{<\rho(z)>}{<\rho(0)>}=\frac{\xi(0)^4}{\xi(z)^4},
\end{equation}
and
\begin{equation}\label{eq:den_er}
\delta\rho=\frac{4<\rho(0)>\xi(0)^4\delta\xi}{\xi(z)^5}.
\end{equation}
The variation in amplitude of the wave can be seen to be a proxy for the variation in density. 

{Before continuing, we note that the equations given in this section were derived assuming that the plasma in the waveguide
is static, i.e. there are no flows along the waveguide. Spicules by their very nature are jets, hence, there will be plasma flow along the waveguide
at the time of the measurements. At present, there is no extension to the magneto-seismological tools shown here that includes
the influence of flows. However, insights into the magnitude of the influence of the flow on the measured parameters can be obtained from, e.g., 
\cite{NAKROB1995}, \cite{TERetal2003}, \cite{MORERD2009b}, \cite{RUD2010, RUD2011}, \cite{MORetal2011b}, \cite{SOLetal2011b}. For example, the 
equations in \cite{SOLetal2011b} suggest that the phase velocity is modified by terms on the order of $(U/c_k)^2$ (e.g., their Eqs.~9 and 11). From 
considering the governing wave equations with flows incorporated, the terms introduced by the flow are also on the order of  $(U/c_k)^2$ or 
$U/c_k^2$.}

{Flows along various spicules in the current data set were found to have an average value of $\sim28$~km/s (\citealp{MOR2012}). In 
Fig.~\ref{fig:flow_obs}, we show a time-distance diagram generated from the unsharp-masked data. The cross-cut is this time placed parallel to the 
axis of the spicule of interest here. This reveals the pattern of flows along the spicule. At the time the wave occurs (approximately 300-500~s in 
Fig.~\ref{fig:flow_obs}), the measurable flow speeds are between 20-30~km/s. Hence, this means that the affect due to the flow on the wave is small, 
$(U/c_k)^2 <0.05$, and can be neglected for now. An advanced theory of wave propagation with magneto-seismological tools that incorporates the 
influence of flow will be required to confirm the validity of this assumption. Furthermore, such a theoretical framework will be highly desirable for future 
studies of chromospheric waves.}
  
\begin{figure*}[!tp]
\centering
\includegraphics[scale=0.6, clip=true, viewport=0.cm 0.0cm 10.cm 9.cm]{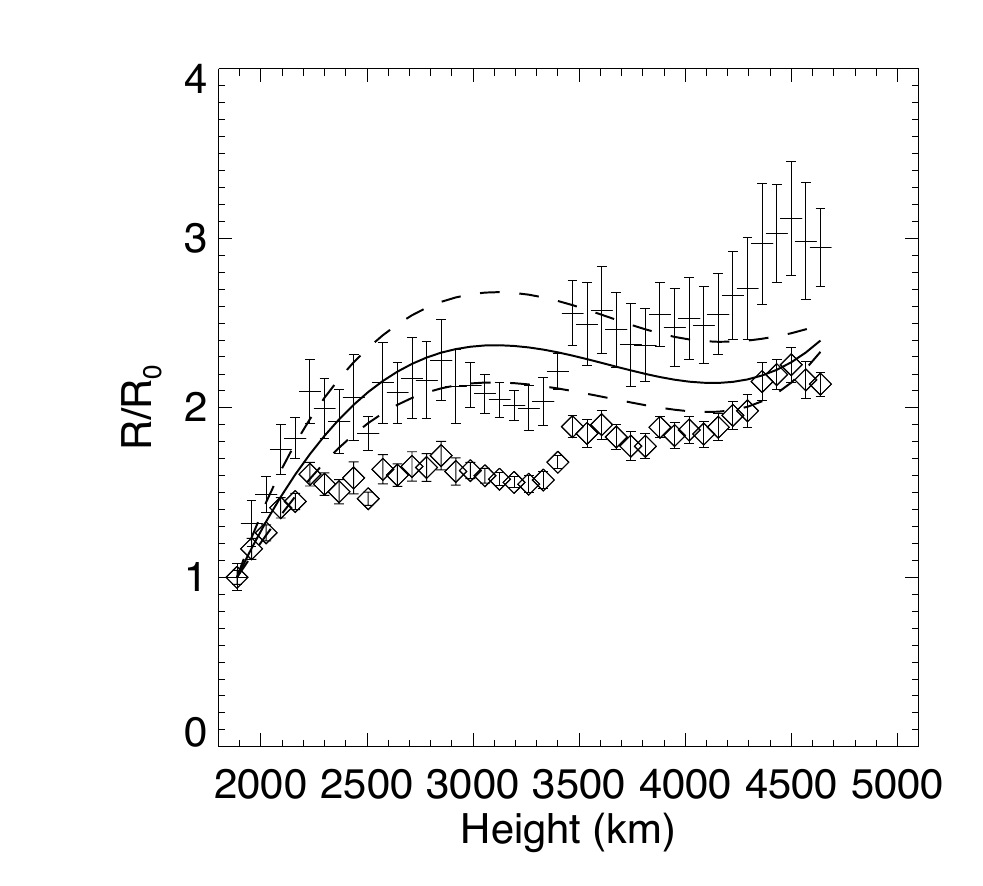} 
\includegraphics[scale=0.6, clip=true, viewport=0.0cm 0.0cm 20.cm 9.cm]{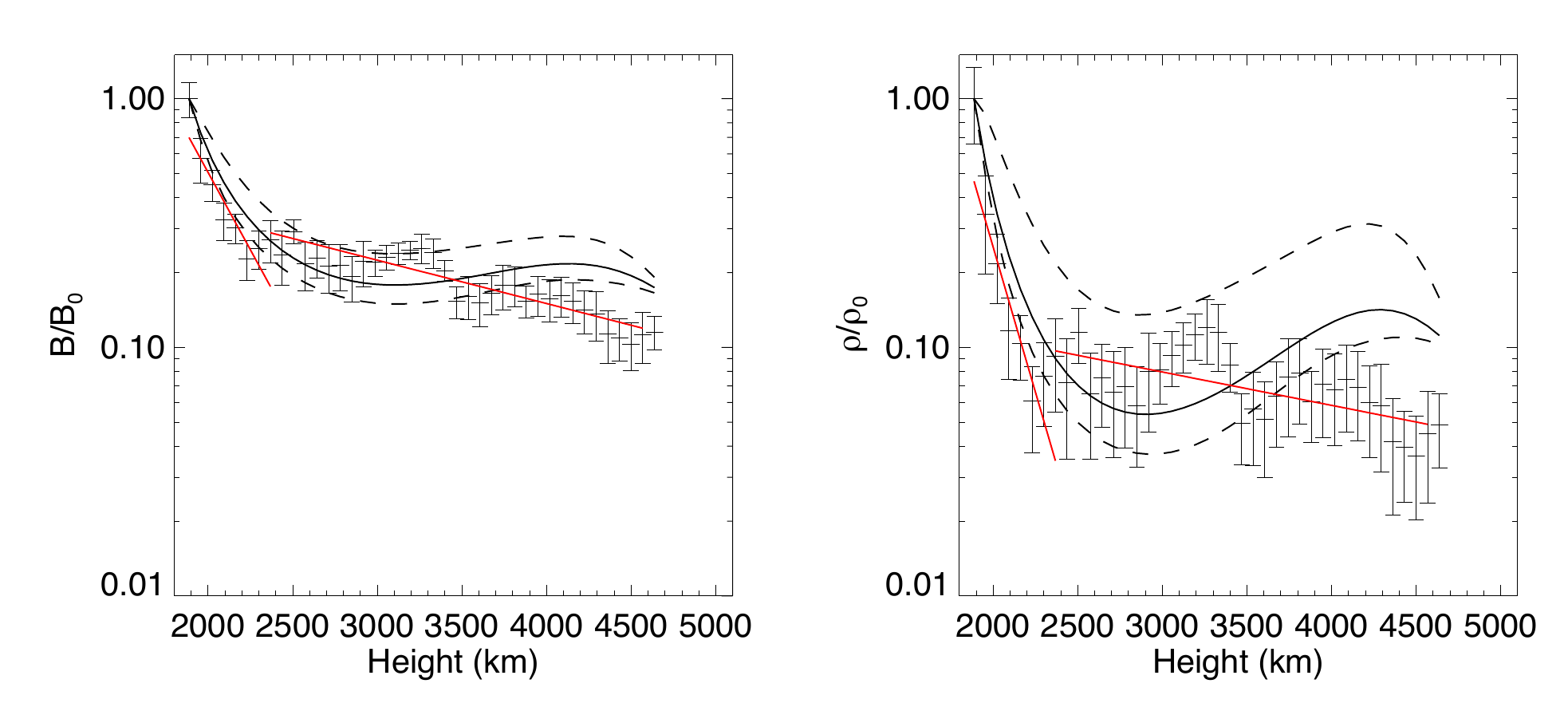} 
\caption{Magneto-seismological estimates of plasma gradients as a function of height in the atmosphere 
relative to $\tau_{5000}=1$. The left hand panel shows the magneto-seismologically determined radial 
expansion (solid line) with the $95\%$ confidence level shown (dashed lines). The diamonds and crosses show the expansion inferred from the 
measurements of the Gaussian width and modelling ({estimated}), respectively, with $\sigma$ errors given. The middle panel shows magneto-
seismological estimate for the gradient of the magnetic field (solid) and the inferred magnetic expansion (crosses). The right hand panel is the density 
gradient determined using the measured displacement amplitude (solid) with $95\%$ confidence intervals (dashed), and a combination of the modelled 
({estimated}) width and measured phase speed (crosses). The red lines are exponential fits to the data points.}\label{fig:sms}
\end{figure*}

\subsection{Intensity measurements}\label{sec:inten}
\begin{figure*}[!tp]
\centering
\includegraphics[scale=0.8, clip=true, viewport=0.cm 0.0cm 10.cm 9.cm]{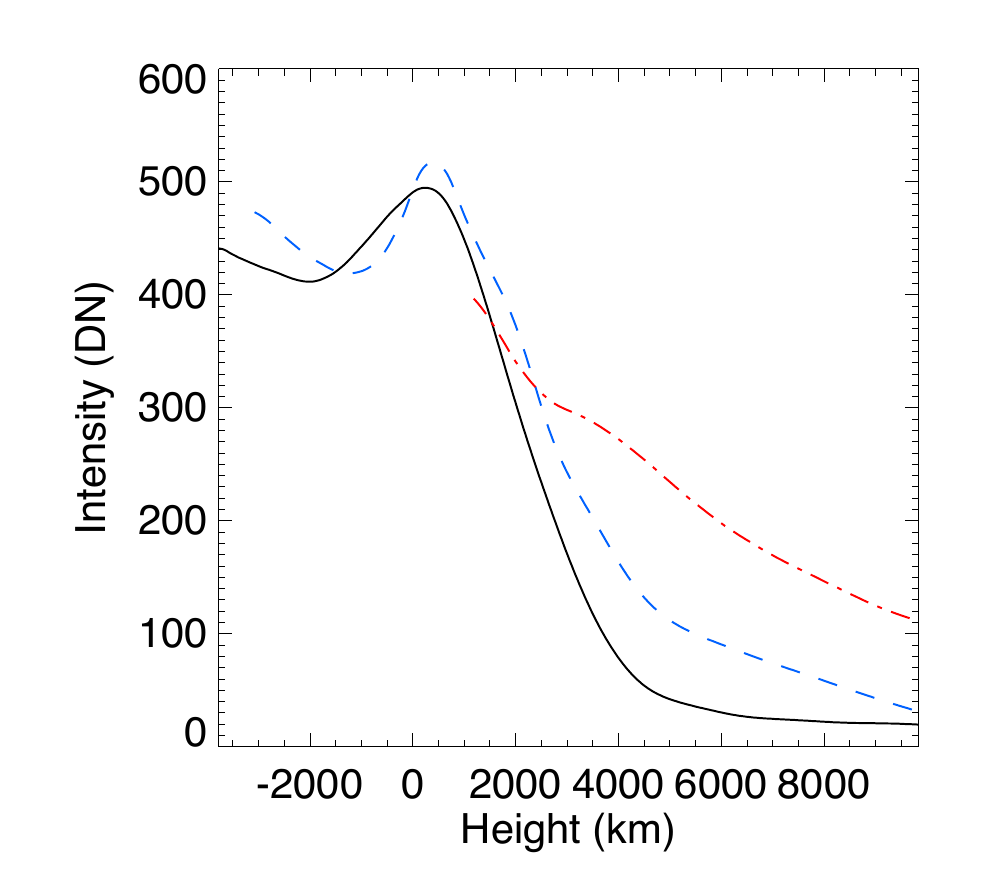} 
\includegraphics[scale=0.8, clip=true, viewport=0.cm 0.0cm 10.cm 9.cm]{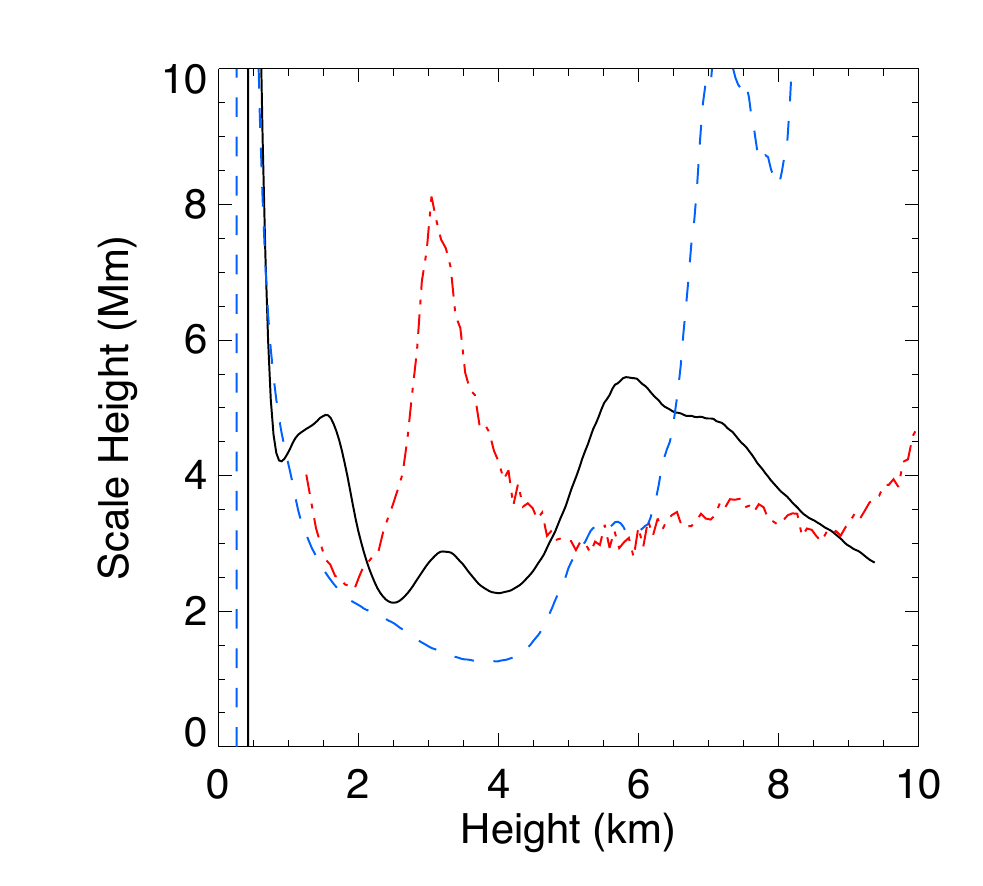} 
\caption{Spatially and temporally averaged intensity profiles (left panel) for the penumbra (black solid), the plage (blue dash) and the spicule (red dash-
dot) as a function of height in the atmosphere relative to $\tau_{5000}=1$. The right panel shows the calculated intensity scale heights for each 
feature.}\label{fig:inten}
\end{figure*}
As well as incorporating the width measurements of the spicule for comparison to magneto-seismological estimates of expansion and magnetic field 
gradients, an estimate of the density scale heights is achievable via other means. Previous studies suggest that the emission gradient can act a guide to 
the density gradient along the spicule (e.g. \citealp{MAK2003}; \citealp{BJO2008}; \citealp{JUDCAR2010}). The assumption is made that the spicule 
plasma can treated essentially as an optically thin media, meaning the intensity is proportional to the density and the source function. Assuming a
constant source function means the intensity scale height equals the density scale height. In the calculations of \cite{JUDCAR2010}, the mean spicule 
source function, calculated locally by collisional excitation and scattering of radiation from the photosphere, decreases by less than a factor of 2 over 
3~Mm. Hence, the measured intensity scale height can only be used as a guide rather than a definitive measurement of the plasma density scale height.

First, we determine the intensity along the spicule by placing a cross-cut parallel to its axis and averaging the values over the time period for 
which the oscillation is measured. In addition, it is of interest to see how the intensity changes as a 
function of height for the penumbra as well as the neighbouring plage region. To measure the intensity scale heights for the surrounding atmosphere, 
cross-cuts are taken at an angle to the limb and averaged over a relatively 
narrow spatial window but an extended temporal window. The intensity scale heights are calculated using
\begin{equation}
H(z)=-\frac{1}{\frac{d}{dz}\ln(I(z))}.
\end{equation}
As this involves taking derivatives of the intensity, discontinuities lead to large variations in the scale height. To suppress these effects, a smoothing of 
the intensity profiles is performed. The intensities and scale heights for the different features are shown in Fig.~\ref{fig:inten}.

Using the pointing information from the observations log, the polar coordinate values of each pixel location are determined with respect to disk centre. 
The angle between the cross-cuts and the normal to the tangent of an arc of constant radius is then obtained, and used later to determine the radial
height of pixels in cross-cuts and correct the values intensity scale heights to give $H(r)$, i.e. the scale height as a function of radial distance. As 
mentioned in Section~\ref{sec:wavefit}, the angle between the spicule and normal to the surface is 28.6 degrees.

For later comparison with existing models of penumbral atmospheres (e.g. \citealp{FONetal2006}), it becomes necessary to determine the exact height 
above the photosphere the radiation is emitted. Following the standard convention, our solar surface (height$=$0~km) is defined as the height at which 
the optical depth equals one for light emitted at 5000~{\AA}. It has been demonstrated by \cite{BJO2008} that the \ion{Ca}{II} formation height at the 
limb is approximately 825~km above this level. In the definition of the \ion{Ca}{II} limb given by \cite{BJO2008}, she identifies the \ion{Ca}{II} limb as the 
point at which the intensity scale height reaches a minimum value, which occurs almost immediately after a steep drop in the value of intensity scale 
height. For the current data, this definition does not give realistic results and would place the limb high into the atmosphere. This is most likely due to 
the apparent closed nature of the \ion{Ca}{II} features seen in the plage regions surrounding the sunspot, which provide an extended emission profile 
above the limb with low variance. To identify the limb, we make use of the visible sunspot umbra. The sunspot has a significantly shortened umbra 
because it has rotated almost off disk and is crossing the limb, no discernible chromospheric emission is obvious on the far side of the sunspot except 
possibly the stalks of some jets. The position of the far-side of the umbra (i.e. the side due to cross the limb first) is taken as the height of the \ion{Ca}
{II} limb.  On relating the intensity profiles with this position, the \ion{Ca}{II} limb occurs close to where the intensity begins to drop off rapidly 
(Fig.~\ref{fig:inten}), after initial intensity bumps from the closed structures in the plage and the bright jets in the penumbral region. For the intensity 
scale heights, it can be seen that the steep drop off in scale height occurs just below the height of 825~km, with the rate of change of scale height 
decreasing significantly above this value. Although the minimum values of scale height do not occur near 825~km, the presence of the steep change in 
scale height below this value provides confidence that the position of the \ion{Ca}{II} limb has been identified reasonably well.

\section{Results}
\subsection{Wave measurements}
In Fig.~\ref{fig:tds}, a number of time-distance diagrams are shown, starting with the lowest altitude cross-cut (top row)
and each of succeeding time-distance diagrams are from successive cross-cuts separated by $800$~km. The periodic transverse 
displacement of the spicule is clearly seen and it is also noticeable that the amplitude of the disturbance increases 
between the first and second rows. The right hand panels show the corresponding fits to the feature in the time-distance diagrams.
It is chosen to fit the wave only for two cycles, additional cycles are evident but the fit result have greater uncertainties.

In Fig.~\ref{fig:observ}, the displacement amplitude measured in all 40 time-distance diagrams is plotted in the first panel. We remind the reader that 
the all measured quantities are given as a function of height in the solar atmosphere above the $\tau_{5000}=1$ height. This is different from the 
distance along the spicule, which can be calculated as $(height -1880)/\cos(28.6)$~km. The amplitude can be seen to 
increase by a factor of two over the first 800~km, then begins to decrease. We speculate in Section~5 as to what is responsible for the observed 
decrease in amplitude and it's implications for the magneto-seismological results. The given amplitude profile is fitted with a cubic polynomial, which 
we note has a smaller $\chi^2_{\nu}$ than a quadratic fit.

\bigskip
In the second and third panels of Fig.~\ref{fig:observ}, the measured lag between time-distance diagrams and the calculated propagation speeds are 
shown, respectively. {The wave is found to be upwardly propagating with no sign of reflection. It has been suggested that a mixture of upward and 
downward propagating waves can exists in coronal hole spicules (\citealp{OKADEP2011}), although the dominant fraction are upwardly propagating. At 
present, we cannot say whether the situation is similar for active region waves.}

The average lag values measured for the $\sim80$~km (black stars) and $\sim40$~km (red stars) separations are given. The $\sim40$~km lag 
values have been multiplied by a factor of two to enable a direct comparison with the $\sim80$~km lag values. It is evident that the two sets of values 
demonstrate a very good agreement. The two sets of lag values are then individually subject to a weighted fit using an exponential function. Linear and 
quadratic functions where also fit to the data points but had greater $\chi^2_\nu$ values than the exponential fit. The fitted function and the calculated 
$\sigma$ uncertainties for the fit parameters are then used to calculate the phase speed and associated error. In Fig.~\ref{fig:observ}, the $\sigma$ 
confidence bound is shown for each of the calculated phase speeds, showing an overlap in values between the two measurements. In the following 
calculations, we will use the results from the $\sim80$~km cross-cut separation as the residuals between the exponential fit and the data points are 
relatively smaller (Section~\ref{sec:phase}).

\subsection{Width measurements}
We now discuss the results of the width measurements before discussing the magneto-seismology results. Fig.~\ref{fig:width_obs} displays the 
measured width of the spicule as a function of distance and measured widths are close to the spatial resolution of \sot (approximately 0.2''=145~km at 
3968~{\AA}). It is clear that the measured values show evidence for expansion along the structure, which to the best of our knowledge has not been 
observed directly along a spicule structure before. The expansion factor reaches a value of $\Gamma=2.2$. Fig.~\ref{fig:width_obs} also displays the 
estimated diameter of the spicule after taking into account instrumental effects discussed in Section~\ref{sec:wid}. The variation in the {estimated} width 
of the structure can be seen to greater ($\Gamma=3$) than that of the half-width measured from the Gaussian fit. It is unclear 
how the use of an idealised PSF compared to a fully calibrated PSF (e.g. \citealp{WED2008}) would influence the current results. The calculated width 
should be taken as a guide to the actual expansion. 

\subsection{Magneto-seismology}\label{sec:mag_seis}
The magneto-seismological results are obtained by exploiting the measured displacement amplitude and phase speed from the 80~km cross-cut 
separation. Instead of the data points, the fitted functions to $\xi$ and $c_k$ are used as inputs for the equations. To derive the expansion of the 
spicule, the measurements are substituted into Eq.~(\ref{eq:rad}) and the errors are calculated using Eq.~(\ref{eq:rad_er}). The results are plotted in the 
first panel of Fig.~\ref{fig:sms}, along with the 95\% confidence level. It is found that the spicule expands with height, which corroborates the width 
measurements. Over-plotting the measured and {estimated} expansion, it can be seen that there is a relatively good agreement between the {estimated} 
expansion and the expansion estimate from magneto-seismology. This is especially the case for the first $1000$~km, however, the {estimated} 
expansion appears to increase to a much greater degree over the last $2000$~km. The measured expansion shows much less agreement 
with the magneto-seismological results than the {estimated} expansion.

Next, the gradient of the magnetic field and errors are obtained using Eqs.~(\ref{eq:mag}) and (\ref{eq:mag_er}) and plotted in the middle panel of 
Fig.~\ref{fig:sms}. A initial rapid decrease in magnetic field strength is found and, as with expansion, the gradient flattens out after 700~km. In 
addition, the gradient in the magnetic field obtained from the {estimated} width is over plotted. As expected from comparing the radii 
results, the magneto-seismological and {estimated} results initially show a good agreement and then the results diverge with height. The variation in 
magnetic field strength from the {estimated} results is subject to a two-part exponential fit in order to determine scale-heights. 

The final panel in Fig.~\ref{fig:sms} shows the gradient in the density along the spicules, which is obtained from the fit to the amplitude and 
Eqs.~(\ref{eq:den}) and (\ref{eq:den_er}). It is found that the magneto-seismological estimate suggests that the density decreases with height. The 
gradient of the density displays a strange profile, indicating the density begins to increase with height after 2500~km. We believe this is an artefact of 
our theoretical assumptions in Section~\ref{sec:theory} and will discuss the nature of the increase in Section~\ref{sec:discuss}. 
{Further, we remind readers that the density profile is the average density profile of the internal and external densities. Hence, Figure~5 probably does 
not correspond to the density variation along the spicule alone.}

In advance of this discussion, another method is used to estimate the {average} gradient in density. From Eq.~(\ref{eq:phase}), it is evident that upon 
being able to measure the expansion of the spicule and the phase speed of the wave propagating along the structure, the gradient in the density can be 
sought via inversion of the measured quantities. To this end, the measured phase speed and {estimated} width are used in Eq.~(\ref{eq:phase}) and the 
results are over plotted in Fig~\ref{fig:sms}. The general trend for the density is now found to decrease along the entire length of the spicule, although 
the rate of change decreases significantly after the first few hundred kilometres. This density is subject to a two-part exponential fit in order to 
determine {average} scale-heights.

\subsection{Intensity measurements}
The intensity measurements have been discussed briefly in Section~\ref{sec:inten}, mainly in relation to the location of 
the $\tau_{5000}=1$ height with respect to the \ion{Ca}{II} measurements. In Fig.~\ref{fig:inten}, the left hand 
panel shows the intensity above the limb for a plage region, the penumbral region and also for the spicule 
of interest. The plage emission can be seen to be brighter than the penumbral region and this is evident from 
looking at the emission in Fig.~\ref{fig:fov}. The jets above the penumbra and umbra are sparsely 
populated in comparison to the plage, leading to the lower intensity values. This was noted in \cite{MOR2012} (see Fig.~7 in that 
paper). The peak of the emission close to the \ion{Ca}{II} limb ($\sim800$~km) appears to occur at marginally 
lower heights in the penumbra than for the plage, indicating the possibility of a lower height of the chromosphere in the active region. 
This point will be discussed in further detail in the Section~5.

The spicule emission initially equals the average emission above the penumbra but diverges after 
$\sim1500$~km, and is observed to be much greater than that of the corresponding average plage and 
penumbral atmospheres at larger heights.

On calculating the scale-heights for each region, it is found for all regions the scale height is typically 
greater than 2~Mm which is larger than the predicted scale heights for a gravitationally stratified 
plasma at $0.01-0.02$~MK, i.e. $250-500$~km. These results are typical for the chromosphere as 
seen at the limb and are related to the extended spicular emission (\citealp{BJO2008}; \citealp{PERetal2012}). Under the 
assumption of an optically thin medium, the density should also have similar scale heights.

\begin{figure}[!tp]
\centering
\includegraphics[scale=0.8, clip=true, viewport=0.cm 0.0cm 9.5cm 8.5cm]{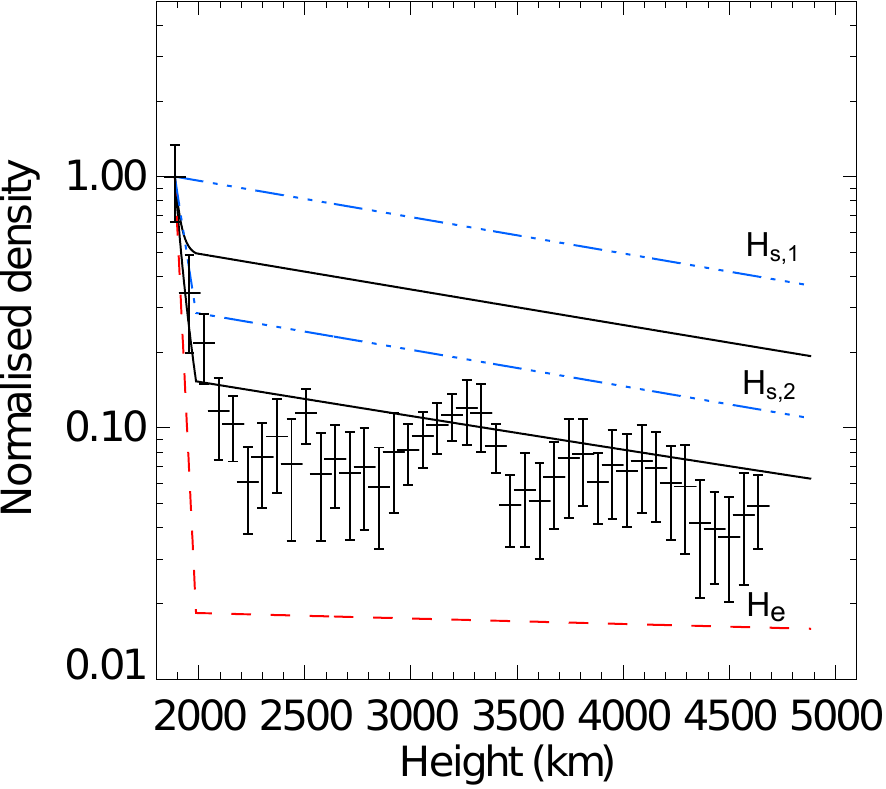} 
\caption{Modelling the predicted density profile along a spicule from magneto-seismology. The data points are the measured change in average density
along the spicule. The red dash line is a normalised density profile ($H_e$) for the ambient plasma, designed as a simplified model of the classic 
hydrostatic solar atmosphere. The blue dash-dot lines show two different normalised internal (spicule) density profiles ($H_{s,1}, H_{s,2}$). The solid 
black lines show the average normalised density profiles calculated using the different internal density profiles. }\label{fig:rho}
\end{figure}

\section{Discussion}\label{sec:discuss}

\subsection{Expansion of the magnetic field}
We begin by discussing the estimate of magnetic field expansion. Both the measured width and magneto-seismological measurements display evidence 
for a rapid expansion over the first few hundred kilometres (atmospheric height of 1800-2500~km). It has been previously noted from 
spectropolarimetric studies above sunspots that the penumbral magnetic fields drops relatively rapidly between the photosphere and chromosphere. 
\cite{RUDetal1995} measures a factor of 2 decrease between photospheric (\ion{Si}{I}) and chromospheric (\ion{He}{I} 10830~{\AA}) magnetic fields 
strengths. Further, \cite{LEKMET2003} measure the majority of magnetic scale-heights to be $<1$~Mm between photospheric (\ion{Fe}{I}) and 
chormospheric (\ion{Na}{I} D1) measurements. Model atmospheres would imply the chromospheric lines used in 
these studies are likely to form in either the low chromosphere, with \ion{Na}{I} formed $<1$~Mm in magnetic concentrations (e.g. 
\citealp{LEEetal2010}) or, upper chromosphere, with \ion{He} {I} 10830~{\AA} showing evidence for the onset of fibrillar structures 
(\citealp{SCHetal2013}) suggesting it samples similar heights to H$\alpha$ and Ca II IR ($\sim1.5-2$~Mm). However, the initial decrease in magnetic 
field strength measured here is more rapid than those reported in these previous studies, requiring a magnetic scale height of $0.35\pm0.05$~Mm. 

\begin{figure}[!tp]
\centering
\includegraphics[scale=0.8, clip=true, viewport=0.cm 0.0cm 10.cm 9.cm]{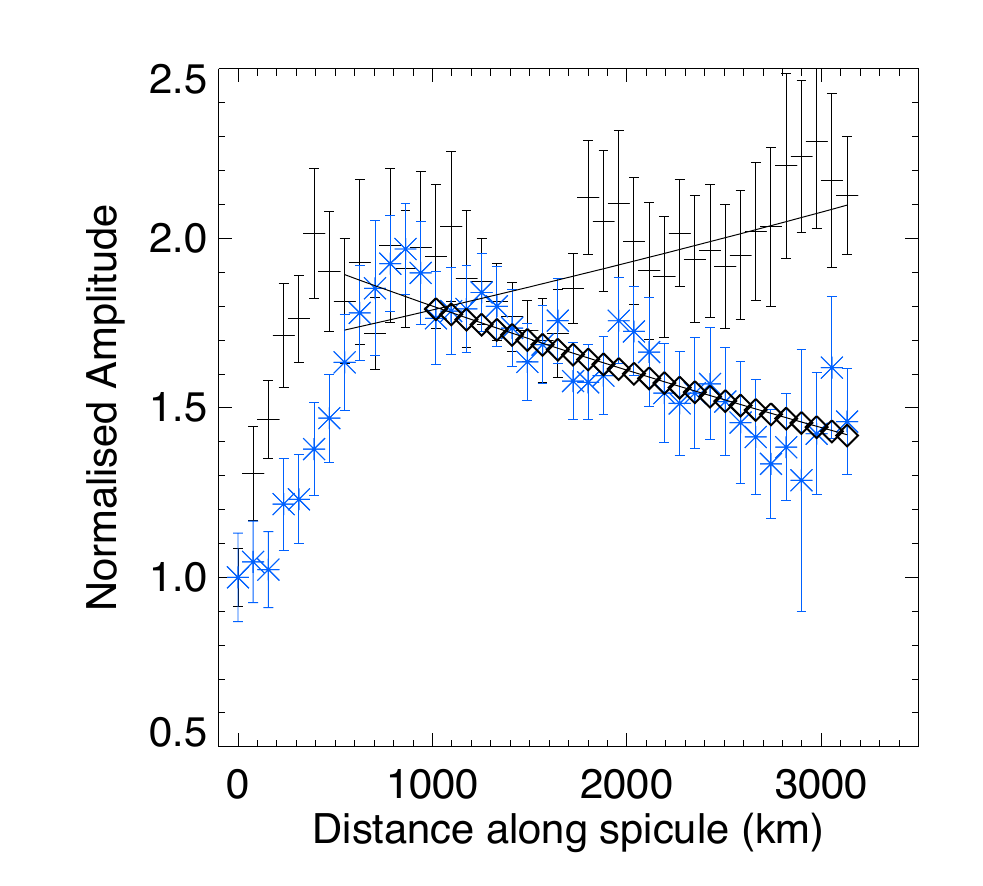} 
\caption{Evidence for wave damping. The plot shows the amplitude measured from the spicule oscillations (blue stars) and the amplitude
determined from the {estimated} expansion and the phase speed measurements (black crosses). The error bars are the $\sigma$ errors for each 
quantity. The black solid lines correspond to exponential fits to each amplitude profile. The diamonds shows the expected amplitude from wave 
damping.}\label{fig:camp}
\end{figure}

The magneto-seismological results show that the spicule then contracts slightly, although 
the error bars do not rule out that the structure remains at a constant radius and magnetic field strength. 
This results is in apparent contradiction to the measurement of the {estimated} spicule width, which 
imply that the expansion after the first 1000~km is much more gradual and has a scale height of 
$2.5\pm0.3$~Mm. The discrepancy between the two results could be due to a variety of reasons. One 
possibility is related to the assumption that the measurement of the width of the spicules cross-sectional flux 
profile provides an estimate of the width of the magnetic structure. The plasma is not necessary optically 
thin at the limb and the emitted radiation is linked to the geometry of the structure. Coupled with this is 
the assumption that the structure expands uniformly in all directions with height. On the other hand, the 
explanation may lie with wave damping. In general, the observed variation in amplitude of a wave will be 
the result of competition between the variation in plasma parameters along the structure, i.e. 
Eq.~(\ref{eq:wkb_res}), and the influence of wave damping due to, e.g. resonant absorption 
(\citealp{TERetal2010c}; \citealp{VERTHetal2010}; \citealp{PASetal2011, PASetal2012}). Considering the 
good agreement between the general trends obtained for the measured {estimated} width and the 
magneto-seismological results during the initial 1000~km, the first option seems less likely as an explanation. 
However, it may be one of the contributing factors to any small scale differences between the two techniques. 
We will return to the discussion of wave damping later and demonstrate that this is a consistent hypothesis for the observed
variation between the two techniques.

\subsection{The density gradient}
Next, we discuss the estimates for the density gradient. The magneto-seismological results 
show similar trends in density to the change in magnetic field, i.e. an large initial decrease in the density 
followed by an increase in density over heights of $3000$~km. Again, we will attribute the increase to wave 
damping. Upon using a combination of the {estimated} width (which is assumed to provide a reasonable 
estimate for the change in magnetic field strength) and the derived phase speed, it is seen that the density 
continues to decrease with height, however at a reduced rate (i.e. implying an increased scale height). {Again, we note that 
the derived variation of density in Figure~\ref{fig:sms}c is an average variation of the internal and external quantities.} The 
estimated {average} scale height for the second phase of density decrease is $3.2\pm1.0$~Mm, which is close
to the measured values of intensity scale height. An exponential fit to the initial phase of density decrease provides
a scale height of $0.19\pm0.04$~Mm, although the over plotted fit in Fig.~\ref{fig:sms} demonstrates an exponential 
does not represent the results adequately. With respect to previous results, the assumption 
of an optically thin plasma is typically made for density estimates from observations in \ion{Ca}{II} (see 
discussion in Section~\ref{sec:inten}). To the best of our knowledge, such observational measurements haven't been 
made for chromospheric plasmas above sunspots and their penumbra

\subsection{A gravitationally stratified background atmosphere}
The presented observations and results raise the question: Can the observed initial rapid decrease in both magnetic field 
strength and density be explained by a physical model? We begin by re-emphasising that the seismological changes in 
both density and magnetic field actually represent the average changes in each quantity (see Eq.~\ref{eq:phase}). The relative 
agreement between the spicule width measurements and the magneto-seismological results for the first $\sim500$~km 
suggest that the internal magnetic field is subject to a rapid decrease. By proxy, we assume that the 
density decrease obtained from the magneto-seismology is also physical. The implied initial scale height is then 
somewhat at odds with that derived from the Ca II emission (Fig.~\ref{fig:inten}). However, this difference can be partly reconciled if we hypothesise that 
the ambient plasma which surrounds the spicule is that of an unseen external atmosphere, whose plasma properties demonstrate a variation that 
corresponds closely to a gravitationally stratified atmosphere in hydrostatic equilibrium. \cite{JUDCAR2010} demonstrated that the presence of such an 
atmosphere would not be visible in Hinode Ca II images of the limb. They suggested the presence of the spicular material in the Ca II 
bandpass is due to the spicule plasma being subject to Dopper shifts from combined flows and wave motions. We note that their results were derived 
for plasma in a coronal hole but the general result should still be true for active region spicules.

We provide a simple model to demonstrate how such a scenario could lead to the measured variation in 
density. The density is given by
\begin{equation}
\rho=\rho_0\exp\left(-\frac{z}{H}\right),
\end{equation}
where $H$ is the density scale height. We assume that the spicule originates in the low chromosphere and initially the internal and external 
densities are approximately similar in magnitude, i.e. $\rho_i\approx\rho_e$, and that the external 
medium has a small initial scale height corresponding to the density 
change associated with a Transition Region. A density scale height value of $H_{ex}=25$~km is selected such that the external 
density decreases by a factor of 100 over the Transition Region to produce a typical chromospheric to coronal density 
transition. After 100~km, the density scale height in the external medium 
is assumed to have a typical low-coronal value of $H_{ex}=20,000$~km (i.e. $T\approx0.4$~MK), i.e.
\begin{equation}
H_{ex}=\left\{\begin{array}{c l}
25~\mbox{km} & 0\le z < 100~\mbox{km},\\
20,000~\mbox{km} & 100\le z < 2800~\mbox{km}.
\end{array}\right.
\end{equation}
Guided by the Ca II emission, we assume that 
spicule material has a constant scale height of $H_{s,1}=3,000$~km. We normalise the density so the magnitudes of the 
density are not required. The calculated variation in average density is shown in Fig.~\ref{fig:rho}. It is seen that the presence of an unseen external 
hydrostatic atmosphere and a slowly changing spicular density would then provide a result that is 
qualitatively similar to that in Fig.~\ref{fig:sms}. However, the magnitude of the initial decrease can only be realised if 
it is assumed that the spicule density also initially decreases relatively rapidly, as seen if the internal spicule density height 
is 
\begin{equation}
H_{s,2}=\left\{\begin{array}{c l}
80~\mbox{km} & 0\le z < 100~\mbox{km},\\
3,000~\mbox{km} & 100\le z < 2800~\mbox{km}.
\end{array}\right.
\end{equation}

If we are to maintain the assumption of a correspondence between emission scale height and density, then this would imply that an
initial steep intensity gradient is somehow masked in the lower atmosphere. This could well be the case if a number of short, spicule like features
contribute to the emission in the lower atmosphere and mask any drop in intensity. Alternatively, it could be that the observed spicule intensity is 
prescribed by the source functions with little correspondence to the density (\citealp{JUDCAR2010}).

If the presence of a hydrostatically stratified atmosphere is indeed what the magneto-seismological estimates demonstrate, then the results show the 
Transition Region of the external plasma above the penumbra occurs at a height of around 
$1800-2000$~km above the $\tau_{5000}=1$ layer. On comparison to the atmospheric model of \cite{FONetal2006} for 
a penumbral region (model R), the observed height is around $300$~km higher than the predicted height. The Transition 
Region (and chromosphere) are predicted to occur at lower heights above regions of increasing magnetic 
activity, approximately 1500~km. At present, the magnitude of error related to our determination of the position of the \ion{Ca}{II} limb is unclear. This 
is due to the ad-hoc technique that had to be used (see Section~\ref{sec:inten}). 

This result is not at odds with observations in \cite{TIAetal2009} and \cite{MOR2012}, which demonstrate that
the Transition Region above sunspots is higher than the Transition Region above plage regions. However, the conclusions of the
previous results would need to be amended to incorporate the current findings. The spicule-like jets do lift chromospheric and Transition 
Region material higher into the atmosphere than the corresponding plage spicules, as is evidenced by the 
greater heights reached by the penumbral spicules (see, \citealp{MOR2012}). However, this is against the 
background of an ambient, almost `hydrostatic' plasma. The presence of an external `hydrostatic' plasma, along with the reduced opacity above 
sunspots, may help explain the limited self reversals observed in Lyman-$\alpha$ lines above penumbras when compared to quiet Sun profiles 
(\citealp{TIAetal2009b,TIAetal2009}; \citealp{CURetal2008}).

\subsection{Evidence for wave damping}
Now, we consider whether the discrepancy between the measured spicule width and the magneto-seismological results can be explained
by wave damping. Assuming that the measured {estimated} expansion provides the correct change in area (and magnetic field strength) along the 
spicule, then, using Eq.~(\ref{eq:rad2}), the expected change in wave amplitude from the variation of the magnetic gradients and phase speed can be 
calculated. The result is plotted in Fig.~\ref{fig:camp} and the height scale on the x-axis now corresponds to distance along the spicule. It can be seen 
that the measured variation in magnetic field strength and phase speed suggest that the amplitude of the wave should increase with height almost 
continuously, in contradiction with the measured amplitude.  On comparing the two amplitude profiles we can then estimate the damping rate in the 
chromosphere. To simplify the situation, we ignore the amplitude measured in the section of the spicule identified with the external Transition Region. 
In addition, it is assumed that the damping length is constant, rather than a function of distance as given in Eq.~(\ref{eq:damp_l}). First, both the 
amplitude profiles are fitted with an exponential profile of the form
$$
A=A_{m/c}\exp\left(\frac{z}{L_{m/c}}\right),
$$
where the subscripts $m$ or $c$ correspond to the measured or calculated values, respectively.  Assuming an exponential damping profile, the rate of 
damping, $L_D$, can be calculated from
$$
\frac{1}{L_m}-\frac{1}{L_c}=-\frac{1}{L_D},
$$
and provides a damping length of $L_D=5500$~km. Multiplying the un-damped amplitude profile by the exponential with the derived damping length 
provides the diamonds in Fig.~\ref{fig:camp}. The quality factor for the damping is calculated to be $\xi=0.34$ for a phase speed of $200$~km/s and 
$P=80$~s.

From these results, it would suggest that there is a relatively strong damping in the spicule feature. The quality factor obtained here is significantly 
smaller than the majority of those previously measured in damped coronal loop oscillations, both for propagating (2.69 - \citealp{VERTHetal2010}) and 
standing ($\xi>1$ - \citealp{VERetal2013b}) modes. 

The quality factor is a frequency dependent quantity, so to compare to previous results in different frequency regimes, it is better to use the quantity 
$\alpha$ introduced in \cite{MORetal2013b}, which is calculated to be $\alpha=L_D/P=0.07$~Mm/s. For estimates of damping between the 
chromosphere and corona, \cite{MORetal2013b} give the value of $\alpha\approx0.2$, which is essentially the value averaged over a height of 15~Mm. 
Considering Eq.~(\ref{eq:damp_l}), it is evident that $\alpha$ will likely increase with height due to the expected trend for the kink (Alfv\'en) speed to 
increase between the chromosphere and corona. Hence the average value in \cite{MORetal2013b} will incorporate a range of $\alpha$ values. The value 
of $\alpha$ we obtain here supports the idea of the enhanced damping of waves propagating through the lower solar atmosphere {when compared to 
wave damping observed in the corona. This is likely to due to a smaller measured phase speed of kink waves in the chromosphere and an increased 
density contrast for chromospheric jets in the corona compared to coronal loops, leading to a smaller quality factor (see, e.g., Eq.~\ref{eq:damp_l})}.

{Note that, at present, other damping mechanisms aside from resonant absorption cannot be ruled out as acting on the observed wave.
For example, in partially ionised plasmas such as the chromosphere, collisions between ions and neutrals can lead to a damping of the waves (e.g.,
\citealp{DEPetal2001}, \citealp{SOLetal2012}). However, the influence of ion-neutral collisions appears negligible compared to resonant absorption for 
the current observation (see, \citealp{SOLetal2012}).}

\section{Conclusions}
Over the last decade, the use of magneto-seismological inversion techniques to probe the solar plasma has increased. However, the ability to test the 
accuracy of results is difficult, mainly because the magneto-seismology is used to estimate otherwise difficult or unmeasurable quantities. The example 
of a spicule oscillation presented here provides a unique opportunity to compare the magneto-seismological results to values measured using 
independent techniques. The comparison between the independent methods shows a good agreement for the variation in spicule plasma parameters 
when the effects of wave damping are negligible compared to variations in plasma quantities. When damping dominates the variation in wave amplitude, 
there is a disagreement between the two independent measures. Hence, this highlights the importance of carefully taking account the various 
competing mechanisms that can affect the amplitude of the wave, in order to ensure that magneto-seismology results provide an accurate description 
of the variations in plasma parameters through the solar atmosphere.   

In carefully measuring wave properties and taking into account the competing mechanisms, we have been able to provide the first direct evidence kink 
wave damping in chromospheric spicules. The derived damping lengths are much shorter than those reported from coronal observations of damped 
kink waves and support the suggestion of an enhanced damping of waves in the lower solar atmosphere (\citealp{MORetal2013b}). The presence of 
enhanced wave damping would also help to explain the strange magneto-seismological predictions for the density variations in \cite{VERTetal2011} and 
\cite{KURetal2013}.

In addition, both direct and magneto-seismological measurement suggest the presence of a strongly stratified atmosphere. On comparing the 
magneto-seismological estimates of density scale height to intensity scale heights in \ion{Ca}{II}, this might suggest a contradiction between the two 
techniques. However, this contradiction can be negated if we assume that there is an unseen strongly stratified ambient plasma in which the weakly 
stratified spicule exists. The ambient atmosphere would be close to the classic hydrostatic model of the solar atmosphere, with a significant density 
gradient in the Transition Region. The notion that an approximately hydrostatic atmosphere that would be unseen in \textit{SOT} \ion{Ca}{II} was initially 
put forward by \cite{JUDCAR2010}. 

Overall, the study presented here highlights the rich vein of information the still remains to be mined from the numerous \hinode and ground based 
images of the chromosphere. While difficulties still remain in more direct measurements of chromospheric plasma properties, the ubiquitous nature of 
MHD kink waves provides an alternative diagnostic option. However, it is worth restating, care is needed when using magneto-seismological techniques 
to diagnose the chromosphere and the assumptions used to derive the diagnostic relations between observables and plasma properties, e.g. 
Eqs.~(\ref{eq:wkb}) and (\ref{eq:phase}), should be always be revisited when interpreting the results of inversions.

\begin{acknowledgements}
RM is grateful to Northumbria University for the award of the Anniversary Fellowship and thanks D. Brooks and G. Verth for 
useful discussions. The author acknowledges IDL support provided by STFC. This work was aided by a 
Royal Astronomical Society Travel Grant.
\end{acknowledgements}

\bibliographystyle{aa}

\end{document}